\documentclass[useAMS,usegraphicx,usenatbib]{mn2e} 
\usepackage{amsmath,graphicx}
\usepackage{subfigure}
\usepackage{ulem}
\usepackage{color}
\usepackage{graphicx}
\usepackage{times} 
\usepackage{amssymb}
\usepackage{amsmath}
\usepackage{lscape}
\usepackage{url}
\newif\ifAMStwofonts
\AMStwofontstrue
\def\rg{${\it r}_{\rm g}$}
\def\rin{${\it r}_{\rm in}$}
\def\laor{\rm{\sc LAOR}}
\def\phabs{\rm{\sc PHABS}}
\def\diskbb{\rm{\sc DISKBB}}

\def\bb{\rm{\sc BB}}
\def\bknpower{\rm{\sc BKNPOWER}}
\def\comptt{\rm{\sc COMPTT}}
\def\po{\rm{\sc PO}}
\def\diskline{\rm{\sc DISKLINE}}

\def\reflionx{\rm{\sc REFLIONX}}

\def\nh{${\it N}_{\rm H}$}
\def\ka{K$\alpha$}
\def\rdblur{\rm{\sc RDBLUR}}

\def\epicmos1{{\it EPIC}{\rm-MOS1}}
\def\epicmos2{{\it EPIC}{\rm-MOS2}}
\def\epicmos{{\it EPIC}{\rm-MOS}}

\def\bepposax{{\it BeppoSAX}}

\def\chandra{{\it Chandra}}

\def\suzaku{{\it SUZAKU}}

\def\xspec{\hbox{\sc XSPEC}}
\def\xspecv{{\sc XSPEC}{\rm\thinspace v\thinspace 12.5.0}}

\def\heasoftv{\hbox{\rm HEASOFT\thinspace v6.6.1}}
\def\xselect{\hbox{\rm XSELECT}}
\def\ftool{\hbox{\rm FTOOL}}

\def\s{\hbox{$\rm\thinspace s$}}
\def\ks{\hbox{$\rm\thinspace ks$}}

\def\deg{$^{\circ}$}  

\def\cm{\hbox{$\rm\thinspace cm$}}

\def\km{{$\rm\thinspace km$}}

\def\ev{\hbox{$\rm\thinspace eV$}}
\def\kev{\hbox{$\rm\thinspace keV$}}

\def\ergpcmsps{\hbox{$\rm\thinspace erg~cm^{-2}~s^{-1}$}}

\def\msun{\hbox{$\rm\thinspace M_{\odot}$}}

\def\n{\hbox{\rm 4U 1705-44}}

\begin{document}

\title[Broad relativistic iron line in \n] {Relativistically broadened
  iron line in the {\it SUZAKU} observation of the neutron star X-ray
  binary \n } \author[Reis, Fabian, \& Young] {\parbox[]{6.in}
  {R.~C.~Reis $^{1}$\thanks{E-mail:
      rcr36@ast.cam.ac.uk},  A.~C.~Fabian$^{1}$ and A.~J.~Young$^2$\\ } \\
  \footnotesize
  $^{1}$Institute of Astronomy, Madingley Road, Cambridge, CB3 0HA\\
  $^{2}$Astrophysics Group, Department of Physics, Bristol University,
  Bristol, BS8 1TL}

\maketitle

\begin{abstract} The X-ray spectra of accreting compact objects often
  exhibit discrete emission features associated with fluorescent
  emission in the accretion disk, the strongest of which is the
  Fe \ka\ fluorescence line at 6.4--6.97\kev. These reflection
  features are amongst the best tools in the study of the inner region
  of accretion flow around a compact object. Here we report on three
  \suzaku\ observations of the neutron star X-ray binary \n\ where a
  broad, skewed Fe \ka\ emission line is clearly visible above the
  continuum.  By using a relativistically-blurred reflection model we
  find that in \n\ the inner disk radius extends down to
  \rin=$10.5^{+1.0}_{-1.7} GM/c^2$ and is at an angle of
  $29.8^{+1.1}_{-1.0}$ degrees to the line of sight. Furthermore, we
  find that the level of ionisation in the surface layers of the
  accretion disk changes by two orders of magnitude between the three
  observations, however the inner radius obtained from the line
  profile remains stable.
\end{abstract}


\section{Introduction}

The X-ray spectra of an accretion disk around a compact object such as
a neutron star often exhibits the signatures of strong gravity. The
most prominent indication of such an extreme environment is the presence
of a relativistically broadened and skewed Fe \ka\ emission line in the
spectra of AGNs (Tanaka et al. 1995; Fabian et al. 1995, 2002;
Brenneman \& Reynolds 2006) and stellar-mass black hole binaries
(Fabian et al. 1989; Miller 2007; Reis et al. 2008, 2009).

The iron-\ka\ fluorescence line arises due to the reprocessing of hard
X-rays in the optically-thick accretion disk (Ross \& Fabian
1993). The shape and degree of broadening of the fluorescence line
gives a direct indication of the extent of the innermost region of
emission. As this region moves closer to the compact object,
gravitational redshift and Doppler effects becomes stronger and
results in pronounced distortion on the shape of the emission line
(Fabian et al. 1989; hereafter we will refer to this as relativistic
broadening). Asymmetric line profiles from neutron star low-mass X-ray
binaries (LMXBs) have recently been reported in a number of sources
(Bhattacharyya \& Strohhmayer 2007; Cackett et al. 2008, 2009; Pandel
et al. 2008; Papitto et al. 2009) confirming the inner accretion disk
origin for these broad lines. Iron fluorescence emission in these
sources can then be used to obtain an upper limit on the radius of the
central neutron star (as it cannot exceed the radius of the inner
accretion disk) and thus constrain the equation of
state of the superdense matter (Lattimer \& Prakash 2007).

Previous observations of the LMXB \n\ have shown the presence of a
broad (FWHM 0.7--1.1\kev) feature at approximately 6.5\kev\ (White et
al. 1986; Barret \& Olive 2002). More recently Di Salvo et al. (2005)
confirmed the presence of this broad line in \n\ using
\chandra. However due to the low effective area above $\sim6.4$\kev\
they were not able to distinguish between relativistic broadening -- as
expected from an accretion disk -- or the broadening expected from
Comptonisation in a hot inner corona. In order to discriminate between
these two scenarios, high statistics are needed to assess the precise
shape of the broad line and in particular the extent of its
red-wing. In this Letter we show that the Fe \ka\ line profiles
obtained in three \suzaku\ observations of \n\ are clearly asymmetric
and compatible with being due to relativistic broadening around the
compact object. By modelling the line profile we constrain the inner
edge of the accretion disk and thus provide an upper limit to the
radius of the neutron star in \n.

\section{Observation and Data reduction}

We observed \n\ on three occasions with \suzaku\ (Mitsuda et al. 2007)
in 2006 August 29 (Obs ID 401046010; hereafter Obs 1), September 18
(401046020; Obs 2) and October 06 (401046030; Obs 3). The four
detectors constituting the X-ray Imaging Spectrometer (XIS; Koyama et
al. 2007) were operated in the burst clock mode for Obs 1 and in the
normal mode for Obs 2 and 3. Obs 1 and 3 were operated in 5x5 and 3x3
editing mode, whereas in Obs 2 the XIS operated in the 3x3 editing
mode only. The burst mode (Obs 1) resulted in a dead-time corrected
exposure of 8.9 and 5.8\ks\ for editing mode 3x3 and 5x5
respectively. This was approximately 80 per cent of the on-source
time. The total exposure time for Obs 2 was 17.3\ks\ for each XIS
camera. Obs 3 resulted in an exposure of 18.7 and 13.8\ks\ for the 3x3
and 5x5 mode respectively.  Using the latest \heasoftv\ software
package we reprocessed the data from the Version 2 processing
following the \suzaku\ Data Reduction
Guide{\footnote{http://heasarc.gsfc.nasa.gov/docs/suzaku/analysis/}}.
Essentially, we started by creating new cleaned event files using the
tool {\rm ``xispi''} and the script {\rm ``xisrepro''} as well as the
associated screening criteria files. \xselect\ was then used to
extract spectral products. We used the script {\rm
  ``xisresp''}{\footnote
  {http://suzaku.gsfc.nasa.gov/docs/suzaku/analysis/xisresp}} with the
``medium'' input to obtain ancillary response files (arfs) and
redistribution matrix files (rmfs). {\rm ``xisresp''} calls the tools
{\rm ``xisrmfgen''} and {\rm ``xissimarfgen''}. Finally, we combined
the spectra and response files from the three front-illuminated
instruments (XIS0, XIS2 and XIS3) using the \ftool\ {\rm
  ``addascaspec''}. This procedure was repeated for each observation
resulting in a total of six XIS spectra. The {\it FTOOL} {\sc grppha}
was used to give at least 20 counts per spectral bin. The Hard X-ray
Detector (HXD; Takahashi et al. 2007) was operated in its normal
mode. The appropriate response and background files for XIS-nominal
pointing were
downloaded{\footnote{http://www.astro.isas.ac.jp/suzaku/analysis/hxd/}}
and the HXD/PIN data were reprocessed in accordance with the \suzaku\
Data Reduction Guide.

We restrict all our XIS analysis to the energy range 1.0--10.0\kev\
and HXD/PIN to 12.0--25.0\kev. Since we are mostly interested in the
iron line profile ($\sim$4.0--7.0\kev) the bulk of the analysis
presented in this Letter focuses on the spectra obtained with the XIS
instrument. All parameters in fits involving different instruments
were tied and a normalisation constant was introduced. \xspecv\
(Arnaud 1996) was used to analyse all spectra. The quoted errors on
the derived model parameters correspond to a 90 per cent confidence
level for one parameter of interest ($\Delta\chi^{2}=2.71$ criterion).

\section{analysis and results}

\begin{figure}
\centering
{
 \rotatebox{0}{
{\includegraphics[height=5.6cm, width=5.5cm]{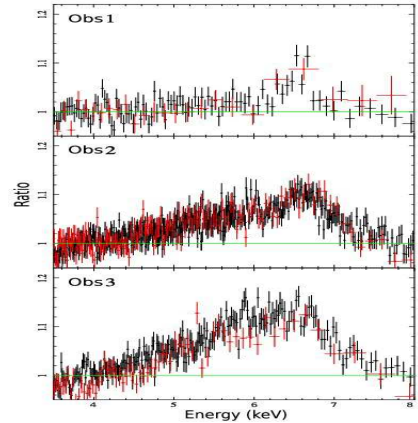}  
}}}
\caption{Data/model ratio emphasising the broad residuals at $\sim
  6.4$\kev\ for \n\ obtained by fitting the energy range 1.0--4.0 and
  7.0--10.0\kev\ with absorbed powerlaw and disk-blackbody. Black and
  red points are for front and back illuminated detectors
  respectively. It is clear that in all three observation there is
  evidence for a broad emission line, however the degree of broadening
  is not constant between them.}

\end{figure}

\begin{table*}
  \caption{Results of fits to \suzaku\ XIS data  }

\begin{tabular}{lccccccccc}                
  \hline
  \hline
  Parameter & &  Model 1       & &&Model 2           &     \\
            & Obs 1 & Obs 2  & Obs 3 & Obs 1& Obs 2& Obs 3  \\

  \nh ($\times10^{22}${\rm cm$^{-2}$})&$2.74\pm0.05$&$2.66\pm0.02$&$2.02\pm0.05$& $2.74\pm0.05$&$2.55\pm0.02$&$1.82\pm0.02$ \\
  $\Gamma$ & $1.83\pm0.01$&$2.12\pm0.01$ & $2.22\pm0.02$ &$1.835\pm0.015$&$2.09\pm0.06$ & $2.01\pm0.03$ \\
  $N_{\rm PL}$ & $0.191\pm0.005$&$1.35\pm0.02$&$0.60\pm0.02$&$0.193\pm0.005$& $1.28\pm0.01$& $0.40^{+0.03}_{-0.02}$\\
  {\it kT} (\kev) & $0.144\pm0.004$ & $0.161\pm0.002$ & $0.189\pm0.015$&$0.143\pm0.004$ & $0.160\pm0.002$ & $0.73^{+0.02}_{-0.03}$ \\
  $N_{\rm MCD}$ ($\times10^3$) & $1730\pm500$ & $2750\pm300$&$38^{+31}_{-20}$ &  $1830^{+580}_{-420}$ & $2210^{+240}_{-220}$&$0.040\pm0.006$\\
  $E_{Gaussian}$ (\kev)& $6.53\pm0.0.05$& $6.40^{+0.002}$ &$6.40^{+0.003}$ &...&...&...\\
  $\sigma$(\kev) &$0.18^{+0.07}_{-0.06}$ &$0.87\pm0.06$&$1.04\pm0.08$&...&...&...\\
  $N_{Gaussian} (\times 10^{-3})$ &$0.232^{+0.068}_{-0.058}$&$5.2\pm0.5$&$3.9\pm0.5$&...&...&... \\
  $E_{\laor}$ (\kev)&... &...&...&$6.64^{+0.06}_{-0.05}$& $6.97_{+0.02}$ &$6.97_{+0.02}$ \\
  $q$ &...&...&... &$2.3^{+0.3}_{-0.4}$ &$3.29^{+0.06}_{-0.05}$& $3.46^{+0.07}_{-0.05}$\\
  $i$ (degrees)  &...&...&...&$17\pm1$ & = Obs 1& =Obs 1\\
  \rin (\rg) &...&...&...&$3.75^{+0.25}_{-0.10}$& = Obs 1& =Obs 1\\
  $N_{\laor} (\times 10^{-3})$ &...&...&...&$0.4\pm0.1$&$5.3^{+0.2}_{-0.3}$&$3.2\pm0.2$ \\
  $EW$ (\ev)& $37^{+11}_{-9}$& $194^{+19}_{-17}$ &  $398^{+56}_{-47}$ & $67^{+21}_{-19}$& $203^{+8}_{-12}$ &  $290^{+17}_{-22}$ && \\
 $\chi^{2}/\nu$&8839.1/8507& = Obs 1& = Obs 1&  8377.5/8505& = Obs 1& = Obs 1\\

\hline
\hline
\end{tabular}

\small Notes.-Model 1 is described in \xspec\ as PHABS$\times$(GA+DISKBB+PL). Model 2 is described in \xspec\ as PHABS$\times$(LAOR+DISKBB+PL). The value of the inclination and \rin\ were tied between the three observations in Model 2. The normalisation of each component is referred to as {\it N}. We note that the value for the inner radius obtained with Model 2 is physically inconsistent  and is only meant to emphasise the need for self-consistent modelling of both relativistic and Compton broadening (see Table 2). 
\end{table*}

The X-ray spectra of accreting neutron stars can generally be
described by a combination of a soft-thermal component such as the
black-body or multicolour-disk blackbody model (\bb\ and \diskbb,
Mitsuda et al. 1984 in \xspec\ respectively), a hard comptonisation
component such as a powerlaw or broken powerlaw (\po\ and \bknpower\
in \xspec\ respectively) or the Comptonisation model by Titarchuk
(1994; \comptt). In addition to this continuum there is also the
presence of a broad emission line at $\sim 6.4$\kev. Lin, Remillard \&
Homan (2007) tested these various combinations on the spectra of two
accreting neutron stars and evaluated their performance against the
$L_X\propto T^4$ criteria. It was found by the authors that a hybrid
model consisting of a black-body, a multicolour-disk blackbody as well
as a powerlaw resulted in the most physically motivated model for
these sources. Anticipating a similar combination for the continuum of
\n, we start by fitting the XIS data in the energy range 1.0--4.0 and
7.0--10.0\kev\ with a powerlaw modified by interstellar absorption
(\phabs{\footnote{{Using the standard BCMC cross-sections
      (Balucinska-Church and McCammon 1992) and ANGR abundances
      (Anders \& Grevesse 1989)}} model in \xspec).  This fit yields
  $\chi^2/\nu= 9928.9/6050$. Adding a \diskbb\ component significantly
  improves the continuum, with $\chi^2/\nu= 6428.5/6044$. The bulk of
  the residuals now comes from the energy range 1.5--2.0\kev\ and is
  probably due to calibration uncertainties. Removing this energy
  range results in $\chi^2/\nu= 4687.4/5216$. Fig. 1 shows the
  data/model ratio for the three observations of \n\ fitted with the
  above model and then extended to the full energy range. It is clear
  from the residuals that a broad emission line is present in all
  three observations of \n, however at a first glance it can be seen
  that the degree of broadening is not constant. We try to quantify
  this by adding a Gaussian line with centroid energy constrained to
  vary between 6.4--6.97\kev, as expected from iron fluorescence
  emission with different ionisation states. The results for this fit
  are shown in Table 1 (Model 1). The strength and broadness of the
  Gaussian line varies significantly between the three
  observations. Notably the equivalent width of the line increases by
  an order of magnitude from Obs 1 ($EW\sim 40$\ev) to Obs 3 ($EW\sim
  400$\ev). To provide a physically motivated description of these
  broad and highly skewed line profiles (see Fig. 1) we replaced the
  Gaussian component with the relativistic model \diskline\ (Fabian et
  al. 1989). This model represents the line emission emerging from an
  accretion disk around a non-rotating, Schwarzchild black hole and
  has been successfully used to model broad emission lines in neutron
  star X-ray binaries (Cackett et al. 2008; Pandel, Kaaret \& Corbel
  2008). The model assumes an emissivity profile described by a
  power-law of the form $\epsilon_{(r)} = r^{\it -q}$ and an inner
  radius \rin\ -- in units of \rg$=GM/c^2$. The outer disc radius was
  fixed at the maximum allowed value of 1000\rg. Only the inner radius
  and disc inclination, {\it i} were tied between the
  observations. This resulted in an improved fit over the Gaussian
  line profile ($\Delta\chi^2= -374$ for 2 degrees of freedom),
  however the inner radius strongly peaks at the minimum value of the
  model (\rin=$6.00^{+0.03}$\rg). We replaced the \diskline\ model
  with that expected around a maximally rotating Kerr black hole
  (\laor, Laor 1991). Table 1 shows the various parameters found for
  this model (Model 2).  The \laor\ line profile results in a good fit
  to the data with $\chi^2/\nu= 8377.5/8505$. The parameters found for
  this model, however, are {\it not} physically consistent. In
  particular, the inner radius obtained with the above model is less
  than the theoretical minimum radius of a neutron star. This result
  was already implied by the peaked value for the inner radius found
  in the \diskline\ fit and is not due to the apparent low disk
  inclination suggested by the model (see Fig. 2).

  The extent of the line broadening is mostly governed by the
  parameter \rin. We can see in this case that the broadening is such
  that it is causing the value of \rin\ to be artificially low. This
  could be a result of broadening due to a {\it combination} of
  relativistic effects (as modelled by \laor\ and \diskline) and
  Comptonisation by a hot, ionised accretion disk. It should be noted
  that the usage of different continuum models does not affect the
  results presented above. To test this we initially fitted the XIS
  data with a continuum comprising  \diskbb\ and the comptonisation
  model \comptt{\footnote{ Using the disk geometry and seed
      temperature tied to that of the MCD component.}}. Similarly to
  the previous continuum, a broad \laor\ line was necessary to model
  the Fe \ka\ residuals having an unphysical inner radius of
  $3.63^{+0.30}_{-0.01}$\rg, and a low inclination. The overall fit
  was, however, worse with $\chi^2/\nu= 8518.1/8501$. Adding the PIN
  data does not change the results presented above.

\begin{figure}
\centering
{
 \rotatebox{270}{
\resizebox{!}{5.cm} 
{\includegraphics{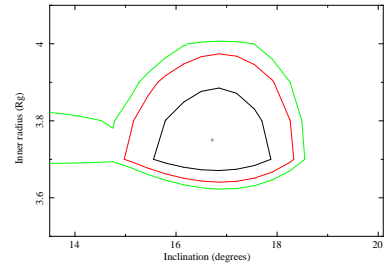}  
}}}
\caption{Inner radius versus inclination contour plot for model
  comprising of disk blackbody, powerlaw and \laor\ emission line
  (Model 2). The 68, 90 and 95 per cent confidence range for two
  parameters of interest are shown in black, red and green
  respectively. It can be seen that the unphysical inner radius
  obtained from this model is not dependent on the value of the
  inclination. This artificially low radius is probably the result of
  a combination of relativistic effects and Comptonisation (see text).
}

\end{figure}

When the Compton-thick accretion disk in such systems are irradiated
by hard X-rays, the ionisation state of the material (especially the
top layers) varies according to the intensity of the hard
X-rays{\footnote{This is the case when the intrinsic thermal
    temperature of the disk is low.}}. The reflected radiation
(including the Fe \ka\ fluorescence line) experiences the effect of
Comptonisation as well as all the relativistic effects mentioned
above. In order to model this reflection self-consistently for the XIS
data, we use the reflection model \reflionx\ (Ross \& Fabian 2005) and
the blurring kernel \rdblur\ which is derived from the code by Fabian
et al. (1989). This combination allows for the interplay between
relativistic and Compton broadening in the total model. The powerlaw
index in \reflionx\ is tied to that of the hard component. The
ionisation parameter $\xi=4\pi F_h/n$, where $F_h$ is the hard X-ray
flux illuminating a disk with a hydrogen density $n$ (Matt, Fabian \&
Ross 1993) is allowed to differ between observations. The model
provides an excellent fit to the various observations of \n\ with
$\chi^2/\nu=8337.6/8505$. Table 2 details the values for the various
parameters in the model, as well as the unabsorbed flux (1--10\kev)
for each model component. The data/model spectra and the best fit
models prior to relativistic blurring are shown in Fig. 3.

\begin{figure}
\centering
{
 \rotatebox{0}{
\resizebox{!}{5.6cm} 
{\includegraphics{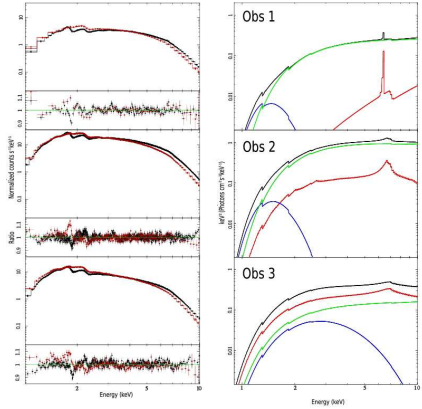}  
}}}
\caption{{\it Left panels:} Data/model ratio for \n\ obtained by
  fitting the XIS data with an absorbed powerlaw, disk-blackbody and
  blurred-reflection. Black and red points are for front and back
  illuminated detectors respectively. {\it Right panel:} Best fit
  model prior to relativistic blurring. The total model, powerlaw,
  disk and reflection components are shown in black, green, blue and
  red respectively. Note that for Obs 2 and 3 the emission line is
  broad prior to relativistic blurring. This is due to the higher
  ionisation state of the accretion disk and the strong effects of
  Comptonisation (see text). }

\end{figure}

From the results presented in Fig. 3 and Table 2 it can be seen that
the ionisation state of the accretion disk strongly influences the
shape of the Fe \ka\ emission line prior to the addition of any
relativistic effects. In the case of Obs 1, the relatively narrow and
peaked emission line (see Fig. 1) is due to minimal effects of disk
Comptonisation. The low colour temperature ($kT\sim0.14$\kev) and
emissivity index ($q\sim3$) indicates that the disk surface suffers
little from strong thermal and radiative ionisation, respectively. The
latter is due to the fact that with such an emissivity profile, the
hard X-rays irradiate a large area of the accretion disk. These
factors combined lead to the low ionisation parameter measured for Obs
1.  The opposite is true for Obs 2 and 3 where the higher disk
colour-temperature and steep emissivity profile{\footnote{A steep
    emissivity profile is interpreted as hard X-ray emission from a
    compact, centrally concentrated region. This intensifies the
    irradiation of the central parts of the accretion disk}} leads to
an increase in the ionisation parameter by over two orders of
magnitude. The effect of thermal radiation from the disk is to keep
the temperature high throughout the surface layers. This causes
further ionisation of elements, particularly those lighter than iron,
and more importantly, results in further Compton-broadening of the
\ka\ emission (Ross \& Fabian 2007). This behaviour dominates in Obs 3
where, due to a much higher disk-surface temperature, the line profile
is particularly broad (see Fig. 3) and the ionisation parameter
high. We note that the reflection model \reflionx\ used does not
include the effects of thermal radiation from deep within the
disk{\footnote{A model which includes such blackbody radiation has
    been developed for stellar mass black hole binaries (Ross \& Fabian
    2007).} however it does, to an extent, account for its effect by
  varying the ionisation parameter accordingly.  The fit to the
  combined observations prior to relativistic blurring resulted in
  $\chi^2/\nu=8675.1/8510$. The addition of \rdblur\ improved the fit
  by $\Delta\chi^2=337$ for 5 degrees of freedom (F-test value = 69).
  The inner accretion-disk radius found here of
  \rin=$10.5^{+1.0}_{-1.7}$\rg\ corresponds to an upper radius of
  $21.8^{+2.1}_{-3.5}$\km\ for a neutron star with a mass of
  1.4\msun. Allowing the inner radius to differ between the various
  observations {\it does not} change the above result as all inner
  radii obtained remains within error of the global value.

\begin{table}
  \caption{Results of fits with model including both relativistic and Compton broadening of the Fe \ka\ line. }
\begin{tabular}{lcccccccccc}                
\hline
\hline
\hline
Parameter & Obs 1 & Obs 2& Obs 3  \\
\hline
\nh\ $(10^{22} \cm^{-2})$&$2.79^{+0.04}_{-0.05}$&$2.61\pm0.02$&$1.84\pm0.01$ \\
$\Gamma$ & $1.88\pm0.02$ & $2.080^{+0.005}_{-0.007}$& $1.92^{+0.04}_{-0.05}$\\
$R_{\rm PL}$ &$0.203\pm0.007$ &$1.12^{+0.02}_{-0.03}$ &$0.16^{+0.06}_{-0.09}$\\
{\it q} & $3.0\pm1$ & $9^{+1}_{-4}$ & $5.0^{+1.1}_{-1.7}$ \\
\rin (\rg) & $10.5^{+1.0}_{-1.7}$&= Obs 1& = Obs 1 \\
{\it i} (\rm deg) & $29.8^{+1.1}_{-1.0}$&= Obs 1& = Obs 1  \\
$\xi ({\rm erg}\cm\s^{-1})$ & $10^{+4}$& $1600^{+500}_{-300}$ & $3570^{+930}_{-750}$ \\
$R_{\rm REFLIONX}^a$ & $174^{+70}_{-47}$ & $2.39^{+0.56}_{-0.48}$ & $1.1^{+0.3}_{-0.2}$\\
{\it kT} (\kev) &$0.141^{+0.003}_{-0.004}$ & $0.160\pm0.002$ & $0.78\pm0.03$ \\
  $N_{\rm MCD}$ ($\times10^3$)&$2270^{+760}_{-600}$& $2600^{+300}_{-270}$&$0.030\pm0.005$\\
$\chi^{2}/\nu$ & $8337.6/8505$&= Obs 1&= Obs 1\\
  $F_{\rm REFLIONX}$ &$0.21\pm0.05$& $6.0^{+0.8}_{-0.3}$&$8.6^{+2.6}_{-1.7}$\\
  $F_{\rm MCD}$ &$2.7\pm0.3$& $9.63 ^{+0.55}_{-0.39}$& $ 1.64 ^{+0.21}_{-0.24 }   $\\
  $F_{\rm PL}$ &$8.61\pm0.03$& $37.74\pm0.04$&$6.36^{+0.04}_{-0.20}$\\

 \hline
 \hline
\end{tabular}

\small  Notes.-  $^a$In units of $10^{-6}$. The various fluxes refers to the unabsorbed flux in the range 1--10\kev\ expressed in units of $\times10^{-10}$\ergpcmsps.  The model is described in \xspec\ as PHABS$\times$RDBLUR(DISKBB+PO+REFLIONX). The value of the inclination and \rin\ were tied between the three observations, however fits were also performed with the latter restriction removed. The values for the inner radius obtained in the three observations were all within the 90 per cent confidence range of the global value.

\end{table}

\section{Discussion}
 
We have observed the low mass X-ray binary \n\ on three different
occasions with \suzaku. The presence of a strong, broad and skewed
iron-\ka\ emission is seen in all observations. The asymmetry of these lines
exclude a Gaussian profile on a statistical basis. The nature of these
broad lines in similar sources have usually been attributed to
relativistic affects and modelled as such using either the \laor\ or
\diskline\ line profile (see e.g. Di Salvo et al. 2005; Cackett et
al. 2009). Similar fits to the \suzaku\ spectra of \n\ results in
unphysical values for the inner radius of the accretion disk. When the
broadening of the Fe \ka\ emission is attributed solely to
relativistic effects the model requires the emission to come from
within 6\rg, less than the theoretical minimum radius of a neutron
star. This artificially low value for the inner radius of emission is
likely due to further Compton broadening in the surface layers of the
accretion disk. Line emission originating in such accretion disks
likely emerges from layers a few Compton depths deep and is thus
Comptonised by the overlying layers. By using a reflection model that
accounts for Compton broadening in the warm, ionised surface layers of
the accretion disk in {\it combination} with relativistic broadening
it was found that the ionisation state of the disk plays an important
part in constraining the innermost emission region for \n.

From Figs. 1 and 3 it can be seen that in our observations of \n\ the
profile of the Fe \ka\ emission changed from being relatively narrow
and peaked to extremely broad. In addition to this, we can see from
Table 1 (Model 2) that the line energy changes from the He-like
6.7\kev\ in Obs 1 to the more ionised H-like 6.97\kev\ in obs 2 and
3. This change is interpreted here as being due to the varying
ionisation state of the disk surface and thus varying levels of
Compton broadening.  The narrow, peaked profile (Obs 1) is typical of
emission from a relatively cold accretion disk with a low ionisation
parameter. As the ionisation parameter ($\xi=4\pi F_h/n$) of the disk
increases -- either from a change in the intensity of the hard X-ray
irradiating the disk or a change in the hydrogen number density --
Compton broadening of the emission line becomes more prominent
(c.f. Obs 2 and 3). From the observed rise in the illuminating flux
and ionisation parameter (Table 2) between Obs 1 and 2 we find that
the hydrogen density in the surface of the disk must decrease by a
factor of $\approx2-10$ between these epochs. The increase in the
illuminating flux between Obs 2 and 3 is, however, enough to account
for the rise in the ionisation parameter without an implicit change in
the hydrogen number density of the disk. A similar study by Ballantyne
\& Strohmayer (2004) has shown that variation in the disk surface
density and ionisation parameter is necessary to model the X-ray
spectra of the neutron star 4U 1820-30.

By simultaneously fitting all three observations and assuming that the
geometry of the disk (inclination and inner radius) stays constant
between them we estimate the inner radius of emission as
\rin=$10.5^{+1.0}_{-1.7}$\rg\ which corresponds to
$15.6^{+1.5}_{-2.5}(M/\msun)$\km . This upper limit on the radius of
the neutron star is in agreement with that presented by Di Salvo et
al. (2005; $<11$\rg) and Piraino et al. (2007; $<12.3$\rg) using
\chandra\ and \bepposax\ respectively.  Timing studies on such objects
can be used as an independent means of measuring the truncation radius
or inner-edge of an accretion disk. The co-rotation radius,
$R_{\Omega} = (GMP^2/4\pi^2)^{1/3}$ specifies the point where a
particle attached to a field line would rotate (magnetosphere
rotation) at the keplerian rate. By assuming that the kilohertz QPO
found in \n\ is associated with the Keplerian frequency at this radius
(1/{\it P}), Olive, Barret \& Gierlinski (2003) found
$R_{\Omega}\sim18(M/\msun)^{1/3}$\km. In order for accretion to take
place in a neutron star LMXB the ratio $R_{\Omega}$/\rin\ must be
close to unity. We find here that this is satisfied for masses ranging
from approximately 1.1 to 1.6\msun. Lattimer \& Prakash (2007) showed
that the maximum neutron star radius, invoking a hard equation of
state, is approximately 14.4 and 14.8\km\ for a neutron star with a
mass of 1.1 and 1.6\msun\ respectively. It thus follows from our
results that there should be a gap of at least $\approx3.5$\km\
between the surface of the neutron star and the edge of the inner
accretion disk in \n\ {\it if} the mass range is restricted to
$>1.4\msun$.

A further product of our study is the measured inclination of the
inner parts of the accretion disk in \n. The value found here of
$29.8^{+1.1}_{-1.0}$ degrees is in agreement with that reported by Piraino
et al. (2007) of 20--50 degrees however it is significantly less than
the value presented by Di Salvo et al. (2005 ; $\sim 60$\deg). We note
that the low value for the inclination found here seems to be
characteristics of neutron stars with broad emission lines. This
potential selection bias has been previously commented on by Cackett
et al. (2008) where for the neutron star LMXBs Serpens X-1, 4U 1820-30
and GX 349+2, they find the inclination to be consistently less than
30 degrees.

\section{Conclusions}

We have studied a set of three \suzaku\ observations of the neutron
star LMXRB \n\ and found evidence of a broad, relativistic Fe \ka\
emission line in their spectra. By using a reflection model in
conjunction with relativistic broadening we measured the inner radius
of the accretion disk and placed an upper limit on the radius of the
neutron star. We have found that the broadening due to Comptonisation
in the warm, ionised surface layers of the accretion disk can result
in an artificially low inner radius. It is thus important to
investigate the interplay between the two broadening mechanism when
using the emission profile to derive physical disk parameters.

The inner radius found here of $21.8^{+2.1}_{-3.5}$\km\, assuming a
neutron star with a mass of 1.4\msun, is in perfect agreement with
that implied by the kHz QPOs in \n. This supports the view that the
broad emission lines seen in the X-ray spectra of neutron star LMXRBs
originates in the innermost regions of an accretion disk.

\section{Acknowledgements}
RCR acknowledges STFC for financial support. ACF thanks the Royal
Society. RCR would like to thank Ed Cackett and Jon Miller for useful
comments and advice on the draft paper and the anonymous referee for
constructive criticism that have greatly improved the paper.

\end{document}